# Comparative Study of Semiconductors for Terahertz Generation by Nonlinear Optical Process: An Overview

Fauzul Rizal[1, 2]

[1]High-field Terahertz Research Group, Szentágothai Research Centre, University of Pécs, Hungary
[2]Institute of Physics, University of Pécs, Hungary

**Email address:**
fauzulr@gamma.ttk.pte.hu (Fauzul Rizal)
[*]Corresponding author

**Abstract:** This study investigates different type of semiconductor crystals including assessing several physical characteristics and relevant parameters, which are refraction index, multiphoton absorption coefficient, electro-optic coefficient, nonlinear coefficient, phase tilting angle, and figure of merit, and mapping them for a comparison to look into options which can be promising alternative efficient sources for Terahertz (THz) generation due to the increasing day to day demand for the efficient sources. The Zinc blende Semiconductor Crystals (ZnTe, ZnSe, GaP, GaAs, and CdTe) as well as newly introduced elemental semiconductors (Selenium and Tellurium) are examined. In this study, we aimed to combine literature review to gather the scattered data about optical properties of semiconductor materials introduced here as well as theoretical calculations. Some of the missing data that is not available from previous research or literature will be calculated and modelled here based on the dielectric function and intrinsic properties of the abovementioned semiconductors. It is discovered that at a higher pumping wavelength, Selenium crystal will make the most promising source for terahertz generation due to its properties compared with the other materials assessed in this study. These freshly introduced semiconductors with different pumping conditions can lead to an optimistic view for the unprecedented efficiencies.

**Keywords:** Terahertz Radiation, Semiconductor, Terahertz Materials, Nonlinear Optics

## 1. Introduction

Terahertz (THz) radiation is an electromagnetic radiation in the frequency range from 0.3 THz to 3 THz, corresponding to wavelengths from 1 mm to 100 μm, although the boundary of these frequency range is not strict and sometimes narrower to only 1 THz. Terahertz band located between microwaves and infrared radiation frequency band, and it shares some properties with each of these other two electromagnetic radiation. THz radiation is non-ionizing radiation (unless on a very high intensity [1]). The THz band had been remained unexplored for a long time due to the lack of useful sources and detectors, making this spectral region often called THz-gap.

Advancement in electronics and photonics after 1970's has provided new materials and devices that made the THz gap accessible and triggered rapid progress in both fundamental research and application. Science and technologies based on Terahertz (THz) frequency electromagnetic radiation have developed rapidly over the last 50 years, to the extent that it recently touches many areas from fundamental science to end-user practical applications [2]. The constantly increasing interest emerging an outlook that one may also consider that THz studies are the frontier area of scientific community itself [3]. The main reason for this rapid development lies on the fact that THz radiation can couple resonantly to numerous fundamental motions of ions, electrons, and electron spins in all phases of matter [4]. Consequently, a wide range of possible implementations of THz technology in various fields, such as imaging, sensing, quality control, wireless communication, biomedicine, and basic science, are gaining a notable attention [5-7]. This ability of THz radiation that can pass through a wide variety of non-conducting materials also makes it become an interesting medium to develop a better imaging technology as well as material characterization [8, 9].

Despite the importance of THz radiation and technology, we lack efficient sources. We have reached the limits of what we can get from Lithium Niobate (LN) crystals (with almost 0.3 % conversion efficiency in cryogenically cooled crystal, 1.4 mJ pulse energy [10] with tilted pulse front pumping technique. Thus, we need alternative sources. There were some promising results with ZnTe with 0.14% conversion efficiency [11], and some interesting calculations with GaP [12], but the scope can be broadened.

This study investigates different type of semiconductor crystals including assessing various physical characteristic and relevant parameters and mapping them for comparison to look for promising alternative efficient sources for Terahertz generation due to the demand for the efficient sources. The zinc blende semiconductor crystals (ZnTe, ZnSe, GaP, GaAs, and CdTe) as well as newly introduced elemental semiconductors (Selenium and Tellurium) are examined. The selenium and tellurium crystalize in helical chain-like structures and thus possess interesting nonlinear optical properties such as huge nonlinear coefficients. This freshly introduced semiconductors with different pumping conditions can lead to optimistic view for unprecedented efficiencies.

## 2. Theoretical Background

### 2.1. THz Generation by Second-Order Nonlinear Optical Processes

When a medium exposed with lights (optical/electric field), the interaction between the optical field and the medium is proportional to the field strength in a linear manner, called linear polarization. However, in nonlinear optics, this polarization is no longer linear. One generally creates nonlinear fields by inducing a nonlinear polarization in the crystal with a strong optical field and measuring the radiation emitted by the decay of the polarization [13]. This strong optical field requirement could easily be provided with an intense short laser pulse. One of the second-order nonlinear optical processes are optical rectification and difference-frequency generation. In terms of THz generation, these two processes generate photon at a THz frequency $\omega_T$ by the interaction of two optical photons at frequencies $\omega_1$ and $\omega_2$ within the nonlinear crystal [14]. THz sources based on second-order nonlinear optical effects difference-frequency generation, optical rectification, and optical parametric oscillation, are being looked up into because of their power scalability [15].

Historically speaking, THz generation (was known as far infrared or sub-mm spectroscopy in the old days) was not easy at the earlier times. Started in 1965, Zernike and Berman performed the first THz generation experiment based on difference-frequency mixing using a ruby laser in a quartz nonlinear optical crystal, placed inside the laser cavity [16]. But limited by technology at the time, characterization of the far infrared output turned to be non-trivial [17]. Later, tuneable THz-frequency output was also demonstrated by difference-frequency generation in other nonlinear optical materials [18]. Both collinear and non-collinear phase matching were used to generate broadly tuneable 0.6-5.7 THz output in LN [19]. In the early 1970s, femtosecond pulsed lasers had not yet been invented, so the experiment had to be carried out with ps laser pulses, and accordingly, the generated THz radiation had a limited bandwidth. Optical rectification of ~20 fs pulses in nonlinear crystals can generate THz radiation with a bandwidth of ~1000 cm$^{-1}$. In 1984, Auston and co-workers [20] showed that a picosecond pulse propagating in a nonlinear medium could induced a moving rectified polarization pulse that would generate Cherenkov THz radiation at a cone angle. The THz radiation appears as fs or sub-ps coherent pulses. Later, collinear phase-matching was achieved in the birefringent crystals GaSe [21], and non-collinear phase matching in the cubic zinc-blende type crystals GaAs [22] and GaP [23]. In GaP, collinear phase matching was also demonstrated [24].

Optical rectification (or electro-optic effects) can also be utilized to generate broadband THz output in a nonlinear optical medium by difference-frequency mixing between pairs of frequency components of the same optical pulse. First demonstrations of this method used picosecond pulses and ZnTe, ZnSe, CdS, and quartz crystals [25] that by the advancement of femtosecond laser, was extended [26], which enabled to generate much broader bandwidths in the THz region. Optical rectification was also demonstrated in zinc-blende type GaAs and GaP crystals [27-29], which have much smaller THz absorption.

### 2.1.1. Terahertz Generation Based on Optical Rectification

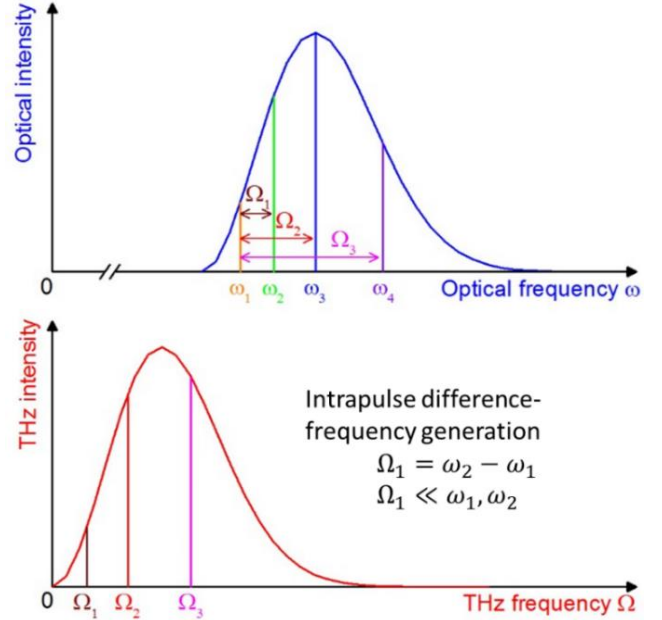

*Figure 1. Difference-frequency mixing among the spectral components inside the pulse. The THz frequency $\Omega_1$ correspond with the difference between optical frequency $\omega_2 - \omega_1$.[4]*

Optical rectification (OR) is an alternative mechanism for pulsed THz generation and is based on the inverse process of the electro-optic effect [30]. Femtosecond laser is required in this process, where the energy of the THz radiation in optical rectification comes directly from the exciting laser pulse. Optical rectification generates photon at a THz frequency $\omega_T$ by the difference-frequency mixing among the spectral components contained within the short pulse (see Figure 1) [14]. Efficient THz generation by OR requires phase matching, i.e., matching the group velocity of the optical pump pulse to the phase velocity of the THz radiation [31]:

$$v^{NIR} = v^{THz} \qquad 1$$

which trivially implied that

$$n_{gr}^{NIR} = n^{THz} \qquad 2$$

where $n_{gr}^{NIR}$ and $n^{THz}$ are index of refraction at near infrared and THz region, respectively. Velocity mismatch between the phase velocity of the THz pulse and the group velocity of the probe pulse leading to a reduction of the electro-optic signal measured on the output, thus distort the THz pulse that we needed as illustrated in figure 2, the THz transient and the intensity envelope of the probe pulse are shown at different times ($t_1 < t_2 < t_3$) while propagating through the crystal. The dashed lines show the intensity envelopes of the probe pulse if there is no velocity mismatch. If there is a velocity mismatch (solid probe pulse envelopes), the probe pulse experiences a different electric field at different positions in the crystal, leading to a reduction of the electro-optic signal.

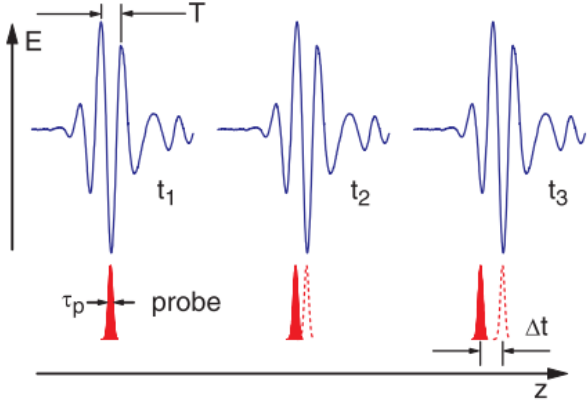

*Figure 2. Effect of a velocity mismatch. [32]*

For phase-matched condition, the efficiency of THz generation by long plane-wave pulses can be described by the following equation [33]:

$$\eta_{THz} = \frac{2\omega^2 d_{eff}^2 L^2 I}{\varepsilon_0 n_{NIR}^2 n_{THz}^2 c^3} \exp\left(-\alpha_{THz}\frac{L}{2}\right) \frac{sinh^2(\alpha_{THz}(L/4))}{(\alpha_{THz}(L/4))^2}. \qquad 3$$

Where $\omega$ is the angular difference (THz) frequency, $d_{eff}$ is the effective nonlinear coefficient, $I$ the intensity of the near-infrared (NIR) light, $\varepsilon_0$ is the vacuum permittivity, $c$ is the speed of light in vacuum, $L$ is the length of the nonlinear crystal, $\alpha_{THz}$ is the absorption coefficient for the THz radiation, $n_{NIR}$ and $n_{THz}$ are the refraction indices in the NIR and THz range, respectively. We clearly see that the conversion efficiency in optical rectification depends primarily on the material's nonlinear coefficient and the phase-matching conditions. For negligible absorption ($\alpha_{THz} L \ll 1$) Eq. 3 will reduce to

$$\eta_{THz} = \frac{2\omega^2 d_{eff}^2 L^2 I}{\varepsilon_0 n_{NIR}^2 n_{THz}^2 c^3} \qquad 4$$

and for large absorption ($\alpha_{THz} L \gg 1$), equation 3 will reduced to

$$\eta_{THz} = \frac{8\omega^2 d_{eff}^2 L^2 I}{\varepsilon_0 n_{NIR}^2 n_{THz}^2 c^3 \alpha_{THz}^2} \qquad 5$$

Optical group indices at several pump wavelengths, the THz refractive indices and absorption coefficients at 1 THz and 3 THz frequency for a few selected materials suitable for optical rectification will be discussed more in detail later in this article.

### 2.1.2. Multiphoton Absorption

In a dielectric material or a semiconductor, linear electronic absorption can occur only if the photon energy is at least as large as the band gap energy. However, at high optical intensity it is possible to bridge that bandgap by simultaneous absorption of two or more lower-energy photons, where the sum of photon energies exceeds the bandgap energy. Such nonlinear absorption processes are called multiphoton absorption (MPA) [34]. The general theory of MPA processes in atomic systems was laid down in the early days of quantum mechanics by Maria Göppert-Mayer, who used nth-order time-dependent perturbation theory to obtain an expression for the probability of the simultaneous absorption of $n$ photons by a single atomic electron. Two-photon absorption was first reported experimentally by Kaiser and Garrett at 1961 [35]. On the intense optical region of semiconductors, the knowledge of multiphoton absorption coefficients is very important. Nonlinear absorption plays a crucial role in limiting the transparency of optical window materials and in causing laser-induced damage to optical components, particularly at short wavelengths [36].

Phenomenologically, MPA can be described by the generalized Beer-Lambert law:

$$\frac{dI}{dx} = -\sum_n \beta_n I^n \qquad 6$$

where $\beta_n$ is the $n$-photon absorption coefficient with [(length)$^{2n-3}$/(power)$^{n-1}$] as the unit, $I$ is the intensity (W/m$^2$), and $x$ is the propagation direction. If for a given field strength and frequency the nth-order process dominates, the attenuation rate of the light flux is

$$\frac{dI}{dt} = -\frac{c}{\sqrt{\varepsilon_\infty}} \alpha_n I^n, \qquad 7$$

where $c$ is the speed of light in vacuum and $\varepsilon_\infty$ is the high-frequency dielectric constant of the material. The relation between $I$ and the peak electric-field amplitude $E_0$ (V m$^{-1}$) as well as the photon number density $N_{ph}$ (m$^{-3}$) is given by

$$\frac{dN_{ph}}{dt} = \frac{\beta_n}{\hbar\omega} I^n = -n\frac{dN_e}{dt}, \qquad 8$$

where $N_e$ is the density of conduction electrons created by the $n$-photon absorption process. In terms of the transition rate $W = dN_e/dt$, $\beta_n$ in equation 6 is given by

$$\beta_n = \frac{2n\hbar\omega(2Z_0)^n W(E_0^{2n})}{\varepsilon_\infty^{n/2} E_0^{2n}}. \qquad 9$$

### 2.2. THz Generation in Semiconductor Crystal

The conversion efficiency in optical rectification depends primarily on the material's nonlinear coefficient and the phase-matching conditions. This technique was first demonstrated for generating far-infrared radiation using LN [25]. Much research has focused on optimizing THz generation through investigating the electro-optic properties of different materials, including semiconductors such as GaAs and ZnTe, and organic crystals [37] among many others.

Because optical rectification relies on coupling of the incident optical power to THz frequencies at low efficiency, it has the advantage of providing very high bandwidths, up to 50 THz [38], although produce lower output powers. Phase-matched optical rectification in GaSe allows ultra-broadband THz pulses to be generated with a tunable center wavelength. Tuning up to a frequency of 41 THz is accomplished by tilting the crystal about the horizontal axis perpendicular to the pump beam to modify the phase-matching conditions [11, 29, 39].

Lithium Niobate (LiNbO3, LN) is the most frequently used intense pulsed THz sources material. It is an artificial ferroelectric crystal that was first produced in Bell Laboratories around 1960's. It is a colorless solid crystal, and transparent for wavelengths between 350 and 5200 nm [40]. It has a non-centrosymmetric trigonal crystal system [41], and possessed nonlinear optical polarizability of various orders, including second-order one.

The large nonlinear coefficient of LN makes it well suitable for optical rectification. However, the THz generation efficiency remains low because of the very different optical group and THz refractive indices [42]. In fact, the optical group index is more than half smaller than the THz refractive index. Therefore, the pump pulse overtakes the THz phase fronts as they propagate through the crystal. Hebling et al. [29] suggested the tilting of the intensity front of the pump pulse to achieve velocity matching in LN. The THz radiation excited by the tilted pulse front of the pump propagates perpendicularly to this front with the speed $v^{THz}$. The angle between the propagation direction of the THz radiation and the pump pulse will be the same as the tilt angle $\gamma$ of the pulse front relative to the phase front of the pump. Then, instead of equation 1, we will have the following equation:

$$v_{gr}^{NIR} \cos \gamma = v_{THz} \qquad 10$$

With the benefit, pulse front tilting of a light beam will also bring angular dispersion:

$$\tan \gamma = -\frac{n}{n_{gr}} \lambda \frac{d\varepsilon}{d\lambda}, \qquad 11$$

with $n_{gr} = n - \lambda \frac{dn}{d\lambda}$ is the group refractive index, and $\frac{d\varepsilon}{d\lambda}$ is the angular dispersion. Larger tilting angle leads to larger angular dispersion. Ravi et al. [43] showed that for one μm pumping, the large frequency down-shift and spectral broadening due to THz generation (cascading effects) accentuates the phase mismatch caused by dispersive effects dominated by group velocity dispersion due to angular dispersion in LN. This turns out to be the strongest limitation to THz generation in LN.

The limitation to THz generation in LN are as following [39]:
(1) Imaging errors in the presence of angular dispersion result in significantly increased pump pulse duration at the edges of a large pump spot [44]. The longer pulse means reduced pump intensity and reduced local pump-to-THz conversion efficiency.
(2) The prism shape of the LN crystal results in THz pulses with different temporal shapes across the THz beam cross section because of the different generation lengths. Such a bad quality, strongly asymmetric THz beam cannot be tightly focused, and thus high THz field strength cannot be achieved.
(3) Due to the large group-delay dispersion (GDD) associated with the angular dispersion (see equation 11), an ultra-short pump pulse evolves very fast inside the LN crystal. Consequently, the average pulse duration becomes much longer than the Fourier transform-limited (FL) pump pulse duration, resulting in strongly reduced effective THz generation length and pump-to-THz conversion efficiency [31].

Several other optimizations for those limitations such as using nonlinear echelon slab [45] and reflective nonlinear slab [46] are already being proposed and calculated to theoretically reach almost 2% conversion efficiency. But we have reached the limits of what we can get from Lithium Niobate crystals (with almost 0.3 % conversion efficiency measured in cryogenically cooled crystal, 1.4 mJ pulse energy) [10] with tilted pulse front pumping technique. Thus, we need alternative sources.

Aside from LN, some early research also shown that several semiconductor materials also able to be the material for generating THz radiation. The earliest findings for measuring optical properties of semiconductor crystals [47-51] between the late 70's to late 90's becomes the base of exploration of the nonlinear interaction in these type of crystal on THz region. The earliest attempt to generate THz radiation (which was called "far-infrared" back then) was done by Yajima and Takeuchi [25] with picosecond laser pulse pump with LN, ZnTe and CdS semiconductor crystal. Not many experimental nor theoretical attempt have been done until the advancement of femtosecond laser technology by Kurtz at the University of Michigan in the early 1990s [52]. Later, collinear phase-matching was achieved in the birefringent crystals GaSe [21], and noncollinear phase matching in the cubic zinc-blende type crystals GaAs [22] and GaP [23]. In GaP, collinear phase matching was also demonstrated [24].

Early experiments on THz generation by OR employed simple geometries, for example: transmission geometry. First realization about the effect of rotating the crystal about its surface normal that makes a systematic variation in the THz emission by OR in 1996 [53] turned the fact that the required phase-matching conditions can be achieved by tilting the pulse-front pump into a certain angle. This and the later [29] study has leads Hargreaves et al. in 2009 [54] to present a cohesive general expression for bulk and surface contributions to OR for planes of arbitrary orientation in zinc-blende crystals. Later, this also opening a gate to carry out several semiconductor materials for generating THz radiation. The potential of semiconductor nonlinear optical materials for high-energy high-field THz pulse generation by OR has been being started to be recognized after these two important findings (experimental and theoretical) are being published.

Limitations of LN as the THz source using tilted-pulse-front pumping, which can pose a serious limitation on increasing the THz yield further has been altering the search of alternative sources. The potential of semiconductor nonlinear optical materials for high-energy high-field THz pulse generation by OR has been recently recognized [55]. At longer pump wavelengths, where in semiconductors usually the optical group velocity is larger than the THz phase velocity, tilted-pulse-front pumping must be used for phase matching. This is where semiconductors play a big advantage over LN. The tilting angle required for semiconductor crystals are typically below 30°, in contrast to LN where it is about 63°. A smaller tilt angle is advantageous for causes a smaller variation of the pump pulse duration within the nonlinear medium and enables a larger effective length for THz generation. This can help to compensate for the smaller nonlinear coefficient of semiconductors. Not only that, but a smaller tilt angle also significantly reduces the spatial inhomogeneity of the interaction length for THz generation. Such a spatial inhomogeneity is a serious drawback in case of LN [43]. The potentially much better spatial homogeneity in case of semiconductor sources enables an easier increase of the pumped area and the THz energy.

Small tilting angle required for phase-matching condition on semiconductor crystal are one of the biggest advantages of using semiconductors as THz materials. Collinear phase matching is even possible in some cases – for example in ZnTe, GaP, and GaAs in the vicinity of 0.8 µm, 1 µm, and 1.5 µm wavelengths. Even though semiconductors – such as ZnTe and GaP – are widely used for THz generation by optical rectification in a low frequency THz range, semiconductors were still considered as less efficient for THz generation than LN. The highest THz energy reported from a semiconductor source was only 1.5 µJ with only $3 \times 10^{-5}$ efficiency [56, 57].

The reason for the low efficiency was the smaller nonlinear coefficient and the strong two-photon absorption at the pump wavelength, associated with the free-carrier absorption at THz frequencies. Two-photon absorption can be avoided, and free carrier absorption can be decreased if we use longer pump wavelengths. Therefore, at longer pump wavelengths, it is possible to suppress low-order multiphoton absorption (by typically requiring tilted pulse-front pumping). As a result, a higher pump intensity can be used, and a higher THz generation efficiency can be expected. Recently, THz pulses with 0.6 µJ energy were generated at $5 \times 10^{-4}$ efficiency by pumping GaAs at a wavelength of 1.8 µm, sufficiently long to suppress 2PA [58]. However, the scope can be broadened, new semiconductors, and other pumping conditions can lead to unprecedented efficiencies.

The potential of long-wavelength pumping of semiconductors for high-energy THz pulse generation has been recognized only recently. Starting from the effort made in 2014 using GaAs crystal pumped at a wavelength of 1.8 µm using a modified tilted-pulse-front scheme, leading to a 0.05% energy conversion efficiency [58]. The spectral content of the measured THz pulses, ranging from 0.1 to 3 THz, are reported to have a good broadband phase matching between the pump and the THz pulses over several millimetres (>20 mm) in a semi-insulating (110) cut bulk GaAs crystal. Those findings also suggest that the pump-to-THz conversion efficiency can be further increased by using a pump source with a narrower bandwidth.

There are some fresh discoveries on ZnTe [56] using pumping beyond the three-photon absorption edge. The resulted THz generation efficiency for optical rectification of femtosecond laser pulses with tilted intensity front in ZnTe was shown to increase 3.5 times compared to pumping below the absorption edge. The four-photon absorption coefficient of ZnTe was estimated to be $\beta_4 = (4 \pm 1) \times 10^{-5}$ cm$^5$/GW$^3$. THz pulses with 14 µJ energy were generated with as high as 0.7% efficiency in ZnTe pumped at 1.7 µm. It is shown that scaling the THz pulse energy to the mJ level by increasing the pump spot size and pump pulse energy is feasible.

The experimental fact that pumping on higher level of MPA could increase the THz efficiency leads to the measurements of three-photon absorption spectra and band gap scaling in direct-gap semiconductors in 2020 [59] which resulting full 3 photon-absorption (3PA) spectra of ZnSe, ZnS, and GaAs as well as the measurements of degenerate 3PA coefficients in InSb, GaAs, CdTe, CdSe, ZnTe, CdS, ZnSe, ZnO, and ZnS. All of them are direct-band semiconductors with zinc-blende structure. This advancement also parallelly followed by the measurement of 4PA in GaP and ZnTe semiconductors in the same year [60].

While the focus for interests is mostly fall on zincblende-type crystal, Shalaby and Hauri [61] have succeed in pioneering to generate an intense THz radiation from trigonal Selenium crystal with broad THz bandwidth (0.5-3.5 THz), high conversion efficiency and THz pulse energy. The spectral THz density produced by optical rectification in Selenium exceeds those from contemporary zincblende-crystal-based THz sources. Few years after that in 2019, Cheng et al. found an interesting property in trigonal Selenium and Tellurium crystal for having large second-harmonic generation and linear electro-optic effect [62]. It was discovered that Tellurium has nonlinear coefficient (d) as huge as 3640 pm/V, while Selenium has electro-optic coefficient (r) as big as 5.63 pm/V. Therefore, trigonal Tellurium and Selenium may find valuable applications for the design of semiconductor photonics devices. This result was later strengthened by further findings from the same team in 2021 [63] which makes these two materials more promising to be a suitable candidate to generate THz radiation by OR.

### 2.3. Data Lacking and How to Determine It

Despite being an important key for alternative THz sources, some of intrinsic properties of most semiconductor crystals are remain unknown. For example, despite the established summary on theoretical models and experimental values of 2PA and 3PA coefficients for a few common materials, little is known about the four-photon absorption (4PA) and nonlinearities of even higher order in semiconductors and other important optical materials. The knowledge of 4PA and higher-order nonlinearities can be crucial for applications driven by infrared pulses. For example, in the newly developed very promising semiconductor THz generators, 4PA can be a major design issue [64]. There is even no report of even 2PA for Selenium and Tellurium.

The most used material to measure MPA coefficients is the z-scan technique. The Z-scan is a simple technique for measuring the change in phase induced on a laser beam upon propagation through a nonlinear material it gives both the sign and magnitude of this phase change, ΔΦ, which is simply related to the change in index of refraction, Δn. Additionally, a Z-scan can also separately determine the change in transmission caused by nonlinear absorption that is related to the change in the absorption coefficient, Δα. Importantly, the determination of the nonlinear refraction from Δn is independent of the determination of the nonlinear absorption from Δα, within a quite broad range of these parameters. Thus, for third-order nonlinear responses the real and imaginary parts of the third-order nonlinear susceptibility, χ(3) can be measured [65].

Beside the higher-order MPA coefficient, the availability of absorption data at THz range are also limited. Terahertz radiation has signatures of very low photon energy (1 THz is equivalent to 4 meV), completely free from damaging to the materials [66]. Nevertheless, in the millimetre-wave (30–300 GHz) and terahertz (0.1–10 THz) frequency bands, EO materials usually have high spreading loss and molecular absorption [67]. Theoretical values from frequency-dependant refractive index function are available for several materials in a broad-spectrum region but with not much details in low-frequency THz region, yet also the experimental data.

Almost all non-centrosymmetric semiconductors exhibit nonlinear optical responses. Although their optical properties have been extensively studied, their nonlinear optical effects have hardly been investigated. For example, Only the second-order nonlinear optical susceptibility element χ(2) has been measured at the CO2 laser frequency for selenium and tellurium and calculated at zero frequency for selenium [62]. No study on the other nonzero element has been reported. Since the OR depends on the electronic band structure, dipole transition matrix and specific frequency and orientation of the applied optical field, it is of interest to know all the nonzero second-harmonic generation elements over the entire optical frequency range, which also related with the need to examine electro-optics effect on them.

With the loads of things to be known to complete the missing information, we felt it would be better if we could collect all the information related to THz wave generation as one complete picture. As this will be useful for mapping in long-term research and useful for helping other researchers who want to enter this field in the future. While trying to calculate the required missing information theoretically based on the available possible information.

# 3. Results and Discussions

## 3.1. Materials of Interest

Nowadays, the most often used materials are ZnTe, GaAs [32], and GaP [68-72]. The advantage of ZnTe is that it fulfils the phase matching condition (group velocity of the pump equal to the phase velocity of the THz radiation) for pump wavelengths around 800 nm and THz frequencies around 1 THz. Thus, ZnTe is suitable for difference frequency generation at up to moderate pump intensities. In contrast, GaAs works very well at high pump intensities, although phase matching is only possible for mid-infrared frequencies. Gallium Arsenide also have a relatively low absorption at the 1 THz frequency. Both ZnTe and GaAs are also usually used in a collinear arrangement makes them easy to align for fine measurement. Gallium Phosphide on the other hand fulfils a nice phase-matching condition for middle-to-higher pump frequencies. With the advancements of interesting numerical calculations [12, 37, 73] and measurement [15] being done recently, this study also considers this material to be compared.

Zinc Selenide, along with ZnTe, GaP, GaAs, and CdTe altogether have a similar zincblende crystal structure. It is also possible to put these materials in a collinear alignment for THz generation. Zinc Selenide also exhibit optical properties similar with ZnTe [74, 75]. Meanwhile, Cadmium Telluride have a similar dispersion behaviour with ZnTe while also exhibit a relatively short band gap energy compared to the other zincblende crystal, thus having a relatively short range of wavelength for MPA. Cadmium Telluride also observed to have a higher nonlinear coefficient [76].

Selenium and Tellurium are birefringence crystals which have a trigonal crystal structure [62, 63, 77], even though Selenium can also be found in an amorphous form as well [78]. These two crystals are farming interest due to their unique optical properties and already observed having a large second-harmonic generation and linear electro-optic effects [62]. Tellurium itself provoking curiosity for expanding the possibilities of using the narrow band gap semiconductor for THz generation due to these findings.

Based on those reasons, we decide to compare the properties of Zinc Telluride (ZnTe), Zinc Selenide (ZnSe), Gallium Phosphide (GaP), Gallium Arsenide (GaAs), Cadmium Telluride (CdTe), Selenium (Se), and Tellurium (Te) as the material to generate THz radiation.

## 3.2. Refractive Index, Group Refractive Index, and THz Refractive Index

This study using pumping wavelength on the range of 800 – 3900 nm, that is: 800 nm, 1030 nm, 2060 nm, and 3900 nm. Some older studies already gathered well-known data for several wavelength range in several catalogue [79, 80], while at the other wavelength inside our range the data scattered in several research. The gap between available data (if any) are being calculated in this study with interpolation and extrapolation or with available Sellmeier equation inside the mentioned wavelength range, in whichever suitable. We calculated the group indexes with the following relations:

$$n_g = n - \lambda \frac{dn}{d\lambda} \qquad 12$$

The well-known dispersion relation for ZnTe has been widely measured and modelled by several studies [79]. The tabulated values from experimental results are being prioritized. While several other values not available from experimental results are being tabulated by calculation of dispersion relation obtained from previous studies [81, 82]. The dispersion relation for ZnSe inside our chosen wavelength range was remodelled from Li et al. [81]. For GaP, the dispersion relation inside our chosen wavelength range was calculated from Madarasz et al. [83]. The dispersion relation for GaAs was remodelled from Hattori et al. [84] for wavelength above 1.2 microns and Skauli et al. [85] for wavelength below 1.2 microns. For CdTe, we remodelled the dispersion relation inside our chosen wavelength range was calculated from Danielewicz [76] and Marple [86]. Since (trigonal) Selenium are birefringence crystal, we consider two dispersion relation inside our chosen wavelength range that was calculated from Kiyoshi and Takuya [87] for ordinary and extraordinary ray. Tellurium is also birefringence crystal as Selenium, but for this case we only focused on ordinary ray configuration that was calculated from Bhar [88].

The THz refractive index at 1 THz for ZnSe is 3.28 [56] and 3.55 for 3 THz [75]. The phase matching condition for this wavelength range are not fulfilled and therefore depends on the angle of pulse-front tilt. The THz refractive index at 1 THz and 3 THz for GaP is 3.34 and 3.37, respectively [23, 79, 89]. The phase matching condition for GaP lies around 1 micron suitable for collinear configuration at short pumping-wavelength. Gallium arsenide inversion zone lies between 0.8 and 1.0 µm indicating a high absorption for this wavelength. For 800 nm pumping wavelength, more energy should be significantly added. The THz refractive index for selenium at 1 THz and 3 THz are 2.80 and 2.82, respectively [61, 90]. Which matched with group index at 1.2- and 1.7-microns pump-wavelength. Unfortunately, that the inversion zone of Te lies inside our chosen wavelength range. With the THz indices at 1 THz are measured as 5.57 [79] and 0.98 at 3 THz (calculated in this work), we can see that no possible matching between THz indices and group indices from 1.2 THz and above. Thus, Tellurium is suitable for THz materials in a limited condition of mid-pumping wavelength with high intensity.

*Table 1. Bandgap energy of several semiconductor crystals.*

| Semiconductors | Bandgap energy (eV) |
|---|---|
| ZnTe [91] | 2.26 |
| ZnSe [89] | 2.67 |
| GaP [56] | 2.27 (indirect) |
|  | 2.48 (direct) |
| GaAs [85] | 1.43 |
| CdTe [89] | 1.43 |
| Se [62] | 2.33 |
| Te [62] | 0.323 |

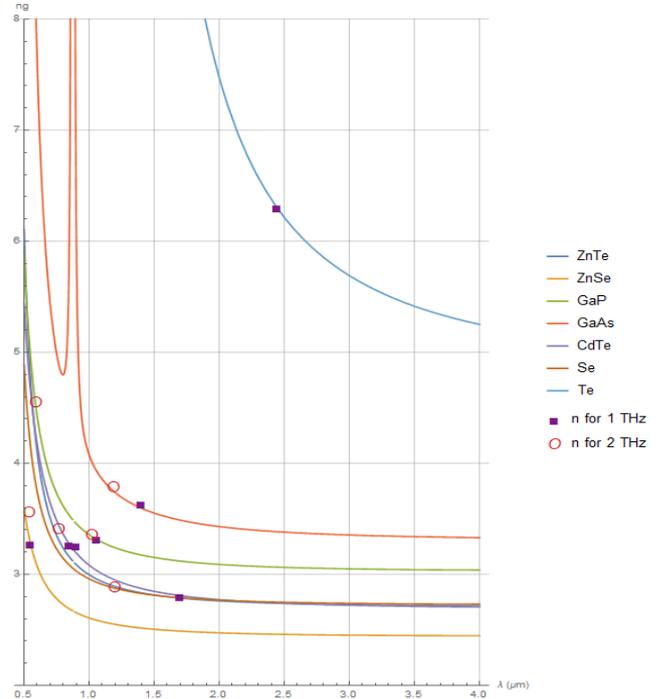

*Figure 3. Chart of group refractive indices for the chosen materials. The square dots and round dots marked the intersection between group index and THz refractive index for each material at 1 and 2 THz, respectively.*

### 3.3. Multiphoton Absorption (MPA)

Multiphoton absorption conditions are directly related with the direct band gap energy of the crystalline solids (Figure 4). The wavelength resemblance of the band gap energy is also the limit of a wavelength range in an order of MPA. Lower band gap means broader range of wavelength on MPA. Take for example, ZnTe with 2.26 eV band gap that correspond to 464 nm wavelength, means that 2PA lies between 464 – 928 nm ranges. So that we know with 800 nm pump wavelength, the effective MPA are 2PA and all that above. The list of band gap energy for selected materials are shown in Table 1.

Two-photon absorption, three-photon absorption (3PA), and the optical Kerr effect (related to the non- linear refractive index, n2) can severely limit performance of THz-wave source based on optical rectification of femtosecond pulses at the high optical intensities required for its operation because of losses due to free-carrier generation, induced free-carrier absorption (at long wavelengths), self-phase modulation, and self-focusing. So, the higher order of MPA that can we achieve with applicable pumping wavelength would be the best option.

### 3.3.1. ZnTe

Valence electrons in Zinc Telluride with 2.26 eV direct band gap can be excited with minimum photon wavelength of 548.6 nm, in which we can effectively pump until 8PA with our chosen 3900 nm pump wavelength. Consequently, for pump wavelengths above 1.65 μm, the absorption edge is for 3PA, only 4PA and higher order MPA are effective.

But unfortunately, the MPA coefficient experimental measurement results for this material are only available until 4PA, which is $3.5 \times 10^{-4}$ cm$^5$/GW$^3$ measured at 1750 nm. This value still corresponds to the same order of MPA with 2060 nm wavelength, so we also include it into data collection. For lower pump wavelength, ZnTe can be used in collinear alignment but with the higher MPA coefficient for 2PA makes the output efficiency would be small as the free-carrier absorption would be higher. The efficiency of THz generation in ZnTe could be increased by two orders of magnitude, from $3.1 \times 10^{-5}$ to as high as 0.7% [60]. The detailed corresponding values of MPA for ZnTe are presented in table 2 below:

*Table 2. MPA coefficients for corresponding chosen pump wavelength for ZnTe.*

| Wavelength (nm) | Wavelength of MPA (nm), lower limit | Order of MPA | MPA coefficient |
|---|---|---|---|
| 800 | 0 | 2PA | 4.2 cm/GW |
| 1030 | 548.6 | 2PA | 4.2 cm/GW [59] |
| 2060* | 1644 | 4PA | $3.5 \times 10^{-4}$ cm$^5$/GW$^3$ [60] |
| 3900 | 3836 | 8PA | ? |

*Measured at 1750 nm with 30 GWcm$^{-2}$ intensity

### 3.3.2. ZnSe

Zinc Selenide have highest band gap energy compared to the other materials chosen in this study, with the direct band gap of 2.67 eV. Large band gap of ZnTe makes it possible to achieve MPA beyond 4PA due to the smaller lower limit of wavelength of MPA. We can expect to gain even 9PA with 3900 nm pump. But again, the MPA coefficient experimental measurement results for this material are only available until 3PA. The detailed corresponding values of MPA for ZnSe are presented in table 3.

*Table 3. MPA coefficients for corresponding chosen pump wavelength for ZnSe.*

| Wavelength (nm) | Wavelength of MPA (nm), lower limit | Order of MPA | MPA coefficient |
|---|---|---|---|
| 800 | 464.36 | 2PA | 5.8 cm/GW [75] |
| 1030 | 928 | 3PA | 0.015 cm$^3$/GW$^2$ [59] |
| 2060 | 1856 | 5PA | ? |
| 3900 | 3702 | 9PA | ? |

### 3.3.3. GaP

GaP is another semiconductor nonlinear material of high interest for efficient THz generation, with 2.48 eV direct band gap. But no experimental data for higher order PA after 4PA coefficient are available. Thus, there is a clearly perceived lack of knowledge on important material data. The value of MPA coefficient for GaP at 4PA written here, $2.6 \times 10^{-4}$ cm5/GW3 was measured at 1750 nm cm5/GW3 measured at 1750 nm via indirect band gap absorption. The absorption edge is on 3PA as GaP, thus only 3PA and higher order are effective. The choice of the pump wavelength longer than the cut-off for 4PA ensured that no inter-band linear, as well as two- and three-photon absorptions had to be taken into account. GaP also have an indirect band gap mode, that shown to have a slightly different behavior compared to its direct band gap mode [92]. The detailed corresponding values of MPA for GaP are presented in table 4.

*Table 4. MPA coefficients for corresponding chosen pump wavelength for GaP.*

| Wavelength (nm) | Wavelength of MPA (nm), lower limit | Order of MPA | MPA coefficient |
|---|---|---|---|
| 800 | 546.92 (indirect), 500.61 (direct) | 2PA | 0.05 (cm/GW) indirect [93] |
| 1030 | 546.92 (indirect), 1001 (direct) | 2PA (direct) 3PA (indirect) | 0.042 cm$^3$/GW$^2$ direct [89] |
| 1750* | 1640 (indirect) | 5PA (indirect) | $2.6 \times 10^{-4}$ cm$^5$/GW$^3$ indirect [60] |
| 2060 | 2002 (direct) | 4PA (direct) | - |
| 3900 | 3828 (indirect), 3504 (direct) | 8PA | ? |

* Measured at 1750 nm with 30 GWcm$^{-2}$ intensity

### 3.3.4. GaAs

Gallium Arsenide are a low-bandgap material with 1.43 eV band gap energy. Having a large wavelength range of MPA for each order, GaAs also have a high absorption at lower pump wavelength for THz generation with absorption edge at 4PA. Making GaAs only effective pumped at 5PA order or higher. The detailed corresponding values of MPA for GaAs are given in table 5.

Table 5. *MPA coefficients for corresponding chosen pump wavelength for GaAs.*

| Wavelength (nm) | Wavelength of MPA (nm), lower limit | Order of MPA | MPA coefficient |
|---|---|---|---|
| 800 | 0 | - | |
| 1030 | 867.02 | 2PA | 23 cm/GW [94] |
| 2060 | 1734 | 3PA | 0.35 cm$^3$/GW$^2$ [59] |
| 3900 | 3468 | 5PA | ? |

### 3.3.5. CdTe

Cadmium Telluride are not only having a same zincblende crystal structure, but also almost identical direct band gap energy with GaAs which is 1.43 eV. From the Table 6, we can see that CdTe have relatively huge MPA coefficient at mid-pump wavelength (2060 nm in our case) with 1.2 cm$^3$/GW$^2$ at 3PA. So only 4PA or higher order are effective. For lower pump wavelength, the higher MPA coefficient for 2PA makes the output efficiency would be small as the free-carrier absorption would be higher. The other details are shown in the Table 6.

Table 6. *MPA coefficients for corresponding chosen pump wavelength for CdTe.*

| Wavelength (nm) | Wavelength of MPA (nm), lower limit | Order of MPA | MPA coefficient |
|---|---|---|---|
| 800 | 0 | - | |
| 1030 | 867.02 | 2PA | 26 cm/GW [91] |
| 2060 | 1734 | 3PA | 1.2 cm$^3$/GW$^2$ [91] |
| 3900 | 3468 | 5PA | ? |

### 3.3.6. Se

Selenium crystal are rarely used for THz materials in the present time, even though the trigonal Selenium have a direct band gap energy in the same order of most common other zincblende materials (such as ZnTe and GaP). As far as our knowledge, no previous studies have been done to measure the MPA beyond 3PA. Surprisingly enough, the absorption edge is on 2PA for Selenium [95], thus 3PA and higher order are already effective. This crystal also has 5 orders smaller effective MPA coefficient value at 3PA compared to the smallest value that we gathered so far. The data related for MPA are shown in the table 7.

Table 7. *MPA coefficients for corresponding chosen pump wavelength for Se.*

| Wavelength (nm) | Wavelength of MPA (nm), lower limit | Order of MPA | MPA coefficient |
|---|---|---|---|
| 800 | 532.12 | 2PA | 9.5 cm/GW [95] |
| 1030 | 1064 | 3PA | $5.2 \times 10^{-8}$ cm$^3$/GW$^2$ [95] |
| 2060 | 1596 | 4PA | ? |
| 3900 | 3724 | 8PA | ? |

### 3.3.7. Te

Tellurium crystal have a very narrow band gap (0.323 eV) thus required a photon with long wavelength to excite the electrons from the ground state. The corresponding lower limit wavelength of MPA for Te is 2755 nm. This shows that Tellurium is not suitable for THz crystal with short-wavelength pump. The high MPA coefficient at 4PA (200 cm/GW) shows a high free-carrier absorption even on the long-wavelength range. We might expect that effectiveness on or after 4PA but the pumping wavelength corresponding to these conditions are beyond the scope of this study. The data related for MPA in Te crystal are shown in table 8.

Table 8. *MPA coefficients for corresponding chosen pump wavelength for Te.*

| Wavelength (nm) | Wavelength of MPA (nm), lower limit | Order of MPA | MPA coefficient |
|---|---|---|---|
| 800 | 0 | - | - |
| 1030 | 0 | - | - |
| 2060 | 0 | - | - |
| 3900 | 2755 | 2PA | - |
| 6000 | 5510 | 3PA | - |
| 10600 | 8265 | 4PA | 200 cm/GW [96] |

### 3.4. Electro-Optic Coefficient and Nonlinear Coefficient

Optical rectification (OR) can occur in the bulk of a material and has been studied in detail in various crystals. Complete expressions have been given for zincblende crystal faces of arbitrary orientation [97]. The terahertz generation in uniaxial birefringent crystals like LN or Se is similar in principle but differs in the important respect that the polarization of the pump beam rotates as it traverses the crystal. These processes directly related to the nonlinear susceptibility of the materials, which in tensor form have the non-vanishing term $d_{14}$ called nonlinear coefficient (in pm/V). Smaller nonlinear coefficient means that longer effective interaction length for OR are needed for generating THz radiation, compared to the ones with higher nonlinear coefficient.

Another effect which may lead to a distortion of the THz pulse is the dependence of the electro-optic coefficient (r) on the THz frequency. The common description of the electro-optic (EO) effect is in terms of the inverse dielectric tensor $\varepsilon^{-1}$. Under the influence of an electric field, $\varepsilon^{-1}$ changes from its zero-field value $\varepsilon^{-1}(0)$ into $\varepsilon^{-1}(E)$ with the following relation:

$$\varepsilon^{-1}(E) = \varepsilon^{-1}(0) + rE \qquad 13$$

The EO tensor r is a third-rank tensor. Since the inverse dielectric tensor is symmetric ($\varepsilon_{ij}^{-1} = \varepsilon_{ji}^{-1}$), the EO tensor has the property of $r_{ijk} = r_{jik}$. Similar to the second-order order susceptibility, r determined by the crystal symmetry. For example, zincblende semiconductors have only one independent tensor component ($r_{xyz} = r_{yzx} = ...$), which means the components are not equal to zero which have all different indices [32]. For an effective detection, EO coefficient should be high. The clamped electro-optic coefficients r can be used to calculate the nonlinear optical coefficient for optical rectification $d_{eff}$ as:

$$d_{eff} = \frac{n_{gr}^4 r}{4} \qquad 14$$

since directly measured $d_{eff}$ values are not available for few materials. The values of EO coefficients r along with their respective effective nonlinear coefficient d are presented in the table 9.

*Table 9. Eo coefficient r and nonlinear coefficient d of selected materials.*

| Semiconductors | ZnTe [98, 99] | ZnSe [92] | GaP [92] | GaAs [85] | CdTe [80] | Se [62] | Te [62] |
|---|---|---|---|---|---|---|---|
| r (pm/V) | 4.04 | 2.00 | 0.97 | 1.43 | 4.5 | 3.58 | 0.3 |
| d (pm/V) | 68.5 | 20.29* | 24.8 | 63.59* | 84.87* | 72.5 | 84.5 |

Values marked with * are calculated in this study.

As a well-known semiconductor for generating THz sources, there are already plenty of values obtained for ZnTe. While ZnTe shows a high effective nonlinear coefficient (around 68 pm/V), it also has high electro-optic coefficient. Tellurium on the other hand have a remarkably high nonlinear coefficient, but with its small EO coefficient makes it not quite effective for detection. ZnSe and GaP have the lowest value of nonlinear coefficient amongst the other materials chosen here, with GaP also have the tendency to be the most challenging material for EO sampling amongst the other zincblende-type semiconductors here. CdTe have surprisingly large EO coefficients due to its high degree of anisotropy amongst the other zincblende crystals and therefore, it has the highest detection response for small frequencies. However, for frequencies over 1 THz the response is limited by phase velocity mismatch and the phonon resonance. The spectral maximum of the detected THz pulse is around 0.5 THz [37].

Selenium on the other hand exhibit a promising result for being measured [62] to have both high EO coefficient and nonlinear coefficient. The phenomenon is explained by the fact that the two materials are helical chains and thus, possess a high degree of anisotropy. This point of view suggests that CdTe, Se, and Te are excellent materials and may find valuable applications in THz generation based on optical rectification.

### 3.5. Tilting Angle

The THz generation efficiency remains low if we rely only on the collinear alignment, because most of the crystals have different optical group and THz refractive indices as we have seen above. Therefore, the pump pulse overtakes the THz phase fronts as they propagate through the crystal. Thus, the intensity front of the pump pulse should be tilted to achieve velocity matching in LN. The THz radiation excited by the tilted pulse front of the pump propagates perpendicularly to this front with the speed $\upsilon THz$. The angle between the propagation direction of the THz radiation and the pump pulse will be the same as the tilt angle $\gamma$ of the pulse front relative to the phase front of the pump Figure 12 shows the geometrical illustration of tilted pulse front, $L_d$ and $L_{eff}$ are the dispersion length and effective THz generation length, respectively.

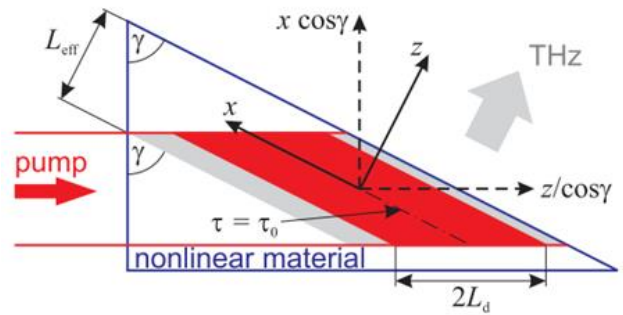

*Figure 4. Chart of group refractive indices for the chosen materials. The square dots and round dots marked the intersection between group index and THz refractive index for each material at 1 and 2 THz, respectively. [31]*

The relation of group velocity and the tilting angle are shown in equation 10. However, pulse front tilting of a light beam will involve angular dispersion as mentioned earlier in the equation 11. Which means with larger tilting angle comes larger angular dispersion. Large angular dispersion brings the large frequency down-shift and spectral broadening due to THz generation (cascading effects) accentuates the phase mismatch caused by dispersive effects dominated by group velocity dispersion. Thus, we want to look for the materials with smallest tilting angle requirements as possible. Table 10 shows the required tilting angle obtained from the previous experiments and from the calculation being done in this study.

*Table 10. Required tilting angle of selected materials for velocity-matching condition.*
*The smallest values for each material are written in bold. The smallest value overall is written in bold underlined.*

| Semiconductors | Wave- length (nm) | Tilting angle (°) | Semiconductors | Wavelength (nm) | Tilting angle (°) |
|---|---|---|---|---|---|
| ZnTe [11] | 800 | - | CdTe | 800 | 19,8 * |
|  | 1030 | **18** |  | 1030 | 28,3 * |
|  | 2060 | 28 |  | 2060 | 11,8 * |
|  | 3900 | 30 |  | 3900 | **6,6** * |
| ZnSe | 800 | 23,7 * | Se (o) | 800 | 25,4 * |
|  | 1030 | 17,9 * |  | 1030 | 35,2 * |
|  | 2060 | **7,0** * |  | 2060 | 13,8 * |
|  | 3900 | × * |  | 3900 (e) | **7,1** * |
| GaP [11] | 800 | - | Te (o) | 800 | - |
|  | 1030 | **6,64** |  | 1030 | - * |
|  | 2060 | 22,53 |  | 2060 | - * |
|  | 3900 | 24,46 |  | 3900 | **45** |
| GaAs [37] | 800 | 31,2 * | Values marked with * are calculated with equation 10. | | |
|  | 1030 | 25,1 * | × = mismatch, - = collinear | | |
|  | 2060 | **14** | (o) = ordinary, (e) = extraordinary | | |
|  | 3900 | 20 | | | |

Our findings shows that the required tilting angle for the chosen materials are in general lies below 45 degrees, far below the ones required for LN (68°). Some of the materials like ZnSe, CdTe, Se, and GaAs shows a downtrend for higher pump wavelength. Zinc selenide have calculated to have a phase mismatch at the 3900 nm pump wavelength. Ordinary ray of Selenium calculated to have a good result for lower pump wavelength except for 3900 nm which shows an incompatibility condition as with ZnSe, but extraordinary ray of Selenium on the other hand shows a surprisingly small tilting angle requirement at this pump wavelength despite being calculated as incompatible for any other chosen wavelength. This behavior also related to the wavelength of 2PA edges. In general, the required tilting angle becomes larger near the 2PA edges. This may be caused by the high induced free-carrier absorption observed around 2PA edges.

### 3.6. Figure of Merit

To compare the suitability of different nonlinear optical crystals for optical rectification, we can start from the formula on the equation 3 which leads to equation 4 for negligible absorption $(\alpha_{THz}L \ll 1)$, or equation 5 for large absorption $(\alpha_{THz}L \gg 1)$. According to those two equations, Hebling et al. [100] introduced two figures of merit (*FOM*) of the nonlinear crystal used for optical rectification defined by:

$$FOM \equiv \frac{d_{eff}^2 L^2}{n_{NIR}^2 n_{THz}}, \qquad 15$$

and

$$FOM_\alpha \equiv \frac{4 d_{eff}^2}{n_{NIR}^2 n_{THz} \alpha_{THz}^2}, \qquad 16$$

which are applicable for weak- and strong-absorbing crystals, respectively. The energy conversion efficiency of optical rectification in the varied materials is proportional to these FOMs. We should notice that for approximately 100 fs pump pulses the crystal length must be limited to 1–2 mm because of dispersion in the THz range [101]. Because of this, and in order to get a unique parameter, it is useful to define the FOM as *FOM* if $\alpha_{THz}$ is larger than 5 cm$^{-1}$, and as *FOM$_\alpha$* with *L*=1 mm in equation 15 for smaller $\alpha_{THz}$ values. We already gathered the values for absorption in 1 THz and 3 THz for the chosen materials in the table 11.

*Table 11. Absorption of materials at 1 THz and 3 THz. The smallest values for corresponding THz frequencies are underlined.*
*Value marked with * are calculated in this study.*

| Semiconductors | ZnTe | ZnSe | GaP | GaAs | CdTe | Se | Te |
|---|---|---|---|---|---|---|---|
| α at 1 THz (1/cm) | 0,32 | 2,2 * | 1,5 | <u>0,13</u> | 2,51 | 4 | 1,11 * |
| α at 3 THz (1/cm) | 11,88 | 6,87 * | 5 | 2,06 | 3,237 | <u>2</u> | 4,12 * |

From table 12, we can see that for most zincblende crystal except CdTe, the FOM getting smaller showing that the energy conversion for 3 THz frequency is getting less efficient compared to FOM for 1 THz. This is where crystal with higher degree of anisotropy such as CdTe, Se, and Te are superior. With Tellurium has the biggest FOM, followed by Selenium and CdTe, at 3 THz. However even if CdTe has the third highest FOM value at 3 THz, this material is strongly absorbing at the 800 nm output of the Ti:sapphire laser, the most commonly used ultrashort pulse source. ZnSe and GaP on the other hand have the smallest FOM compared to the other materials presented here. GaP already reported to generate THz frequency with 0.05% conversion efficiency with high-energy optical parametric amplifier and 144 fs pump pulse duration [11].

*Table 12. Calculated figure-of-merit (FOM) for selected materials at 1 THz and 3 THz.*
*The highest values for each corresponding THz frequencies are marked in bold underlined. (o) = Ordinary ray, (e) = Extraordinary ray*

| Semiconductors | Wavelength (nm) | FOM at 1 THz (pm$^2$/V$^2$) | FOM at 3 THz (pm$^2$/V$^2$) |
|---|---|---|---|
| ZnTe | 800 | 177,94 | 43,72 |
|  | 1030 | 186,96 | 51,79 |
|  | 2060 | 196,12 | 60,28 |
|  | 3900 | 198,39 | 62,12 |
| ZnSe | 800 | 19,71 | 1,29 |
|  | 1030 | 20,32 | 1,44 |
|  | 2060 | 20,99 | 1,62 |
|  | 3900 | 21,20 | 1,66 |
| GaP | 800 | 18,17 | 2,87 |
|  | 1030 | 19,02 | 3,31 |
|  | 2060 | 19,97 | 3,83 |
|  | 3900 | 20,26 | 3,95 |
| GaAs | 800 | 83,99 | 60,66 |
|  | 1030 | 92,07 | 74,49 |
|  | 2060 | 100,66 | 94,30 |
|  | 3900 | 102,63 | 99,47 |
| CdTe | 800 | 257,58 | 161,67 |
|  | 1030 | 294,26 | 161,88 |
|  | 2060 | 304,21 | 206,78 |
|  | 3900 | **311,03** | 217,67 |
| Se (o) | 800 | 232,55 | 187,68 |
|  | 1030 | 241,16 | 159,34 |
|  | 2060 | 257,18 | 240,03 |
|  | 3900(e) | 157,12 | 153,06 |
| Te (o) | 800 | 66,54 | 305,90 |
|  | 1030 | 6,31 | 309,50 |
|  | 2060 | 42,90 | **312,07** |
|  | 3900 | 52,61 | 304,85 |

## 3.7. Comparison

From the data that has been presented in this study, we can see that even though ZnTe and GaP is the most popular semiconductors for THz generation, there are several other materials that are more superior in several aspect. Small FOM of GaP and ZnSe are a major drawback to setup a highly efficient THz sources. Furthermore, the high absorption at higher THz frequency also gives another challenge to consider this material further for THz generation. However, both GaP and ZnTe has a nice velocity matching condition at lower pump wavelength and makes them suitable for collinear configuration.

Gallium arsenide inversion zone lies between 0.8 and 1.0 μm indicating a high absorption for this wavelength. For 800 nm pumping wavelength, more energy should be significantly added. Having a large wavelength range of MPA for each order, GaAs also have a high absorption at lower pump wavelength for THz generation with absorption edge at 4PA. Making GaAs only effective pumped at 5PA order or higher. GaAs also inhibit a large nonlinear coefficient, but a relatively small EO coefficient compared to the others. With its small required-tilting-angle and small absorption at 1 THz as well, we can expect GaAs as the good material for THz generations.

On the other hand, CdTe have surprisingly large EO coefficients due to its high degree of anisotropy amongst the other zincblende crystals and therefore, it has the highest detection response for small frequencies. This material also exhibits the highest FOM amongst the other zincblende materials presented in this study and calculated tilting angle below 25°. But as far as our knowledge, we are not aware of any reported use of this material for optical rectification. The reason is that CdSe is strongly absorbing at the 800 nm output of the Ti:sapphire laser, which is the most commonly used ultrashort pulse source. This behaviour of CdTe also limiting its potential to be an efficient crystal for THz generation.

Tellurium despite having huge nonlinear coefficient and low THz absorption at 1 THz (making it have large FOM), are strictly required long pump wavelength and large tilting angle (45°). Due to its narrow bandgap, the corresponding lower limit wavelength of MPA for Te is 2755 nm. The high MPA coefficient at 4PA (20 cm/GW) shows a high free-carrier absorption even on the long-wavelength range.

Selenium crystal have surprisingly huge EO coefficient and nonlinear coefficient as well. Combined with the big FOM values, this material looks very interesting to be considered as the best material candidate for THz generation by OR. Surprisingly enough, the absorption edge is on 2PA for Selenium, thus 3PA and higher order are already effective. This crystal also has 5 orders smaller effective MPA coefficient value at 3PA compared to the smallest value that we gathered so far for the other materials chosen here. The only major drawback of Se is the high required tilting angle value which can reach 35° for 1030 nm pump wavelength. Not to mention that ordinary ray of Se shows disparity of index at 3900 nm pump, even though the extraordinary ray shows a very good velocity matching with only 7° tilting. But despite this drawback, Selenium is still superior from a lot of aspect compared to the other materials chosen here. Some results [61] also showing a spectrally intense, broadband THz source can be built with this crystal.

# 4. Conclusion

This study had already been investigated different type of semiconductor crystals including assessing various physical characteristic and relevant parameters and mapping them for a comparation to look for promising alternative efficient sources for Terahertz (THz) generation due to the demand of the efficient sources. The Zincblende Semiconductor Crystals (ZnTe, ZnSe, GaP, GaAs, and CdTe) as well as newly introduced elemental semiconductors (Selenium and Tellurium) are examined.

By combining literature review to gather the scattered data about optical properties of semiconductor materials introduced here as well as theoretical calculations, some missing data that is not available from previous research or literature are already being calculated and modelled. The comparison is being done by using Figure of Merit (FOM) parameter involving the absorption, nonlinear coefficient and EO coefficient, pulse front tilt angle and order of Multiphoton Absorption (MPA).

This study has shown an interesting result with Selenium crystal that after being calculated, has biggest FOM amongst the others due to the high nonlinear coefficient, combined with high EO coefficient. This result makes this material as a prominent candidate for terahertz crystal, especially at higher pumping wavelengths. Even though the only major drawback of Se is the high required tilting angle value which can reach 35° for 1030 nm pump wavelength. Tellurium however could not be considered as a better choice as an efficient terahertz crystal due to the short bandgap, corresponding to long range of wavelength of MPA, implicating that it has a high free-carrier absorption. ZnSe could be tuned up to 9PA at higher wavelength, but the small nonlinear coefficient makes the efficiency ratio that implied in the FOM relatively as small as the common GaP crystal. More measurements for MPA coefficients for CdTe and Te are needed in the future, as well as for Se.

By the best choice, Selenium crystal will make the most promising source for terahertz generation due to its properties compared with the other materials assessed in this study.